\documentclass[%
reprint,
 amsmath,amssymb,
 aps,
]{revtex4-2}

\usepackage{graphicx}
\usepackage{dcolumn}
\usepackage{bm}
\usepackage{siunitx}

\usepackage{amsmath}
\usepackage{float}
\usepackage{xcolor}
\usepackage{lineno}
\begin{document}

\title{A Universal Constitutive Relation for Static Droplet Retention}

\author{Lisen Bi}
\affiliation{%
 Institute for Applied Materials - Microstructure Modelling and Simulation (IAM-MMS),
Karlsruhe Institute of Technology (KIT), Strasse am Forum 7, 76131 Karlsruhe, Germany}%

\author{Fei Wang}%
 \email{fei.wang@kit.edu}
\affiliation{%
 Institute for Applied Materials - Microstructure Modelling and Simulation (IAM-MMS),
Karlsruhe Institute of Technology (KIT), Strasse am Forum 7, 76131 Karlsruhe, Germany
and Institute of Nanotechnology (INT), Karlsruhe Institute of Technology (KIT),
Hermann-von-Helmholtz-Platz 1, 76344 Eggenstein-Leopoldshafen, Germany 
}%

\author{Ruize Yin}%
\affiliation{%
 Institute for Applied Materials - Microstructure Modelling and Simulation (IAM-MMS),
Karlsruhe Institute of Technology (KIT), Strasse am Forum 7, 76131 Karlsruhe, Germany
and Institute of Nanotechnology (INT), Karlsruhe Institute of Technology (KIT),
Hermann-von-Helmholtz-Platz 1, 76344 Eggenstein-Leopoldshafen, Germany 
}%

\author{Britta Nestler}
\affiliation{
 Institute for Applied Materials - Microstructure Modelling and Simulation (IAM-MMS),
Karlsruhe Institute of Technology (KIT), Strasse am Forum 7, 76131 Karlsruhe, Germany;
Institute of Nanotechnology (INT), Karlsruhe Institute of Technology (KIT),
Hermann-von-Helmholtz-Platz 1, 76344 Eggenstein-Leopoldshafen, Germany;
and Institute of Digital Materials Science, Karlsruhe University of Applied Sciences,
Moltkestrasse 30, 76133 Karlsruhe, Germany
}%

\date{\today}

\begin{abstract}
{Under applied forcing, static droplet retention is accompanied by contact angle asymmetry, with unequal apparent contact angles on opposite sides}. Yet, the relation between the asymmetry and forcing lacks a common constitutive basis. Here, we show that the asymmetry emerges from minimizing the forcing-dependent free energy. The response of the asymmetry to forcing defines a susceptibility whose zero-forcing value yields a {universal} constitutive relation between the asymmetry and forcing. The constitutive relation is quantitatively validated across liquids, volumes, and forcing levels.
\end{abstract}

\maketitle

\section{Introduction}
Droplet retention on solid substrates is widespread in nature, as seen in raindrops clinging to leaves and windows, and is important in technologies including coating~\cite{eral_contact_2013}, printing~\cite{derby_inkjet_2010}, and liquid manipulation~\cite{chaudhury_how_1992}. A basic manifestation of such retention under forcing is static contact angle asymmetry, in which a stationary droplet exhibits different apparent contact angles on its fore and aft sides~\cite{extrand_liquid_1995,quere_drops_1998}. How this asymmetry relates to the applied forcing is therefore central to understanding droplet retention. 

For decades, the forcing--asymmetry relation has been formulated from different physical viewpoints. Three representative formulations are summarized as follows: (i) The Kawasaki--Furmidge model~\cite{kawasaki_study_1960,furmidge_studies_1962} starts from the capillary traction acting along the contact line; integrating its component along the forcing direction yields a force proportional to contact angle asymmetry, with the contact line geometry entering through a {empirical} factor. (ii)The McHale model~\cite{mchale_friction_2022,mchale_adhesive_2025} recasts droplet retention as an Amontons-like friction law, in which the normal load is replaced by a capillary normal force. The applied forcing is then related to this load through an effective friction coefficient determined by contact angle asymmetry. (iii) The Tadmor model~\cite{tadmor_line_2008,tadmor_line_2004} attributes the forcing--asymmetry coupling to interfacial deformation induced by the uncompensated normal component of surface tension. The strength of this coupling is governed by an interfacial modulus. Despite these distinct formulations, whether the forcing--asymmetry relation admits a common constitutive basis remains unresolved.

\begin{figure}[b]
\includegraphics{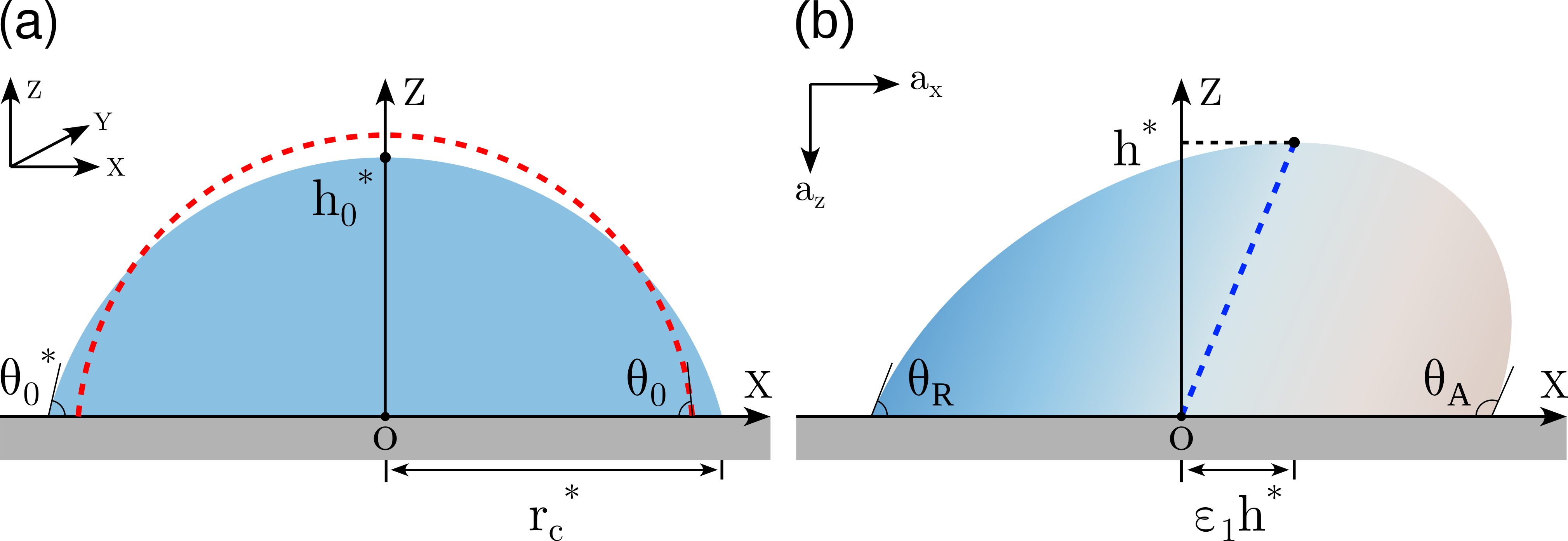}
\caption{\label{Fig.1} Interfacial and bulk effects of the pressure field on a droplet under a uniform body force. (a) Interfacial effect: the pressure-dependent wall free energy changes the zero-forcing configuration (red dashed) to the spherical-cap reference configuration. (b) Bulk effect: the spatially varying bulk pressure field, indicated by the color gradient, deforms the reference configuration into the equilibrium configuration. $\theta_{\mathrm R}$ and $\theta_{\mathrm A}$ are the apparent contact angles on the $-x$ and $+x$ sides, respectively, $h^*$ is the equilibrium height, and $\varepsilon_1^* h^*$ denotes the lateral displacement of the droplet apex.
}
\end{figure}

In this Letter, we resolve this long-standing ambiguity by demonstrating that the forcing--asymmetry relation emerges as a macroscopic equilibrium response governed by the free energy landscape. We incorporate the pressure field generated by the applied forcing consistently into both the interfacial and bulk contributions to the droplet free energy. Minimization over the resulting free energy landscape selects the equilibrium state, from which both apparent contact angles and their asymmetry follow. As the forcing varies, these selected states trace a branch, with the local variation of the asymmetry quantified by a susceptibility. Near the unforced state, we obtain a constitutive relation linking the forcing to the contact angle asymmetry, with the zero-forcing susceptibility as the response coefficient. 

\section{Model}
We consider a liquid droplet of fixed volume $V$ attached to a smooth, homogeneous solid substrate under a uniform body force, see Fig.~\ref{Fig.1}. The free energy functional of the system can be written as 
\begin{equation}
E
=
\int_A \sigma\,\mathrm{d}A
-
\int_S \Delta\gamma(p_s)\,\mathrm{d}S
-
\int_V p\,\mathrm{d}V .
\label{Eq.1}
\end{equation}
The first term is the liquid--gas interfacial free energy, where $A$ is the liquid--gas interface and $\sigma$ is its interfacial tension. The second term accounts for the wall free energy over the wetted region $S$, with $\Delta\gamma\equiv\gamma_G-\gamma_L$, where $\gamma_G$ and $\gamma_L$ are the solid--gas and solid--liquid interfacial tensions, respectively. In conventional macroscopic descriptions, $\Delta\gamma$ is commonly taken as a constant~\cite{whyman_rigorous_2008,marmur_wetting_2003}; motivated by Ref.~\cite{wang_wetting_phenomena_2024,aurbach_microscopic_2026,yin_microscopic_2026}, we retain its dependence on the pressure $p_s$ at the solid--fluid interface and write $\Delta\gamma=\Delta\gamma(p_s)$. Under isothermal local thermodynamic equilibrium, the Gibbs--Duhem relation~\cite{gibbs_1906} links $p_s$ to the bulk pressure field $p=p(\mathbf{r})$, {where $\mathbf{r}=(x,y,z)$ is the position vector}, so that $p_s=p|_S$. The last term depicts the bulk free energy expressed in pressure form~\cite{wang_thermodynamics_2026}. Its origin can be traced to the local volume constraint. For a fluid material element of fixed volume $\mathrm{d}V$, the constraint yields the Lagrange-multiplier contribution $-p\,\mathrm{d}V$~\cite{wang_density_2026}. In the continuum description, integrating these contributions over the entire droplet recovers $-\int_V p\,\mathrm{d}V$.

Equation~\eqref{Eq.1} incorporates the body force through $p(\mathbf r)$. When the droplet experiences a uniform body force, the induced pressure gradient satisfies $\nabla p=\Delta\rho\,\mathbf a$. The integrated form is $p(\mathbf r)=p_0+\Delta\rho\,\mathbf a\cdot\mathbf r$, where $\Delta\rho$ is the liquid--gas density difference, $\mathbf a$ is the body-force acceleration, and $p_0$ is an integration constant setting the pressure at the chosen reference position. It follows from Eq.~\eqref{Eq.1} that $p(\mathbf r)$ affects the droplet in two distinct ways. One is an interfacial effect [Fig.~\ref{Fig.1}(a)] arising through $\Delta\gamma(p_s)$. According to the transversality condition at the droplet edge~\cite{wang_wetting_phenomena_2024,bormashenko_young_2009,blokhuis_young_2009,starov_surface_2009}, the pressure dependence does not alter the form of Young's equation, $\Delta\gamma(p_s)=\sigma\cos\theta_0^{*}$, where $\theta_0^*$ is the corresponding equilibrium contact angle. The interfacial effect can therefore be absorbed into $\theta_0^{*}$. The other is a bulk effect [Fig.~\ref{Fig.1}(b)] arising from $-\int_V p\,\mathrm dV$, whose influence on the droplet geometry is determined by free energy minimization about the reference configuration characterized by $\theta_0^*$.

We first quantify the response of $\theta_0^*$ to the applied forcing. We choose $z$ normal to the substrate and $x$ along the prescribed in-plane forcing direction (Fig.~\ref{Fig.1}), such that $\mathbf a=(a_x,0,a_z)$. Taking $\theta_0$ as the equilibrium contact angle in the absence of forcing, we apply the relation for $\Delta\gamma(p_s)$ in Ref.~\cite{wang_wetting_phenomena_2024} to the present scenario:
\begin{equation}
\cos\theta_0^*-\cos\theta_0
=
\ell\xi\frac{\Delta\rho a_x}{\sigma},
\label{Eq.2}
\end{equation}
where $\ell$ is the adsorption layer thickness~\cite{wang_wetting_2024,aurbach_microscopic_2026,yin_microscopic_2026} and $\xi=V/\pi r_c^{*2}$, with $r_c^*$ the contact radius of the reference configuration[Fig.~\ref{Fig.1}(a)]. Since $\ell$ is not independently accessible, we use the linearized form of Eq.~\eqref{Eq.2} $\theta_0^*=\theta_0-\Lambda L^2\Delta\rho a_x/\sigma$, where
$L=(3V/4\pi)^{1/3}$ and $\Lambda=\ell\xi/(L^2\sin\theta_0)$.
A single water dataset~\cite{berejnov_effect_2007} gives $\Lambda=0.473$, with a
maximum correction of
$|\theta_0^*-\theta_0|\approx 8^\circ$.
The calibrated $\Lambda$ is retained for all other droplet volumes and liquids without further adjustment.

\begin{figure*}
\includegraphics{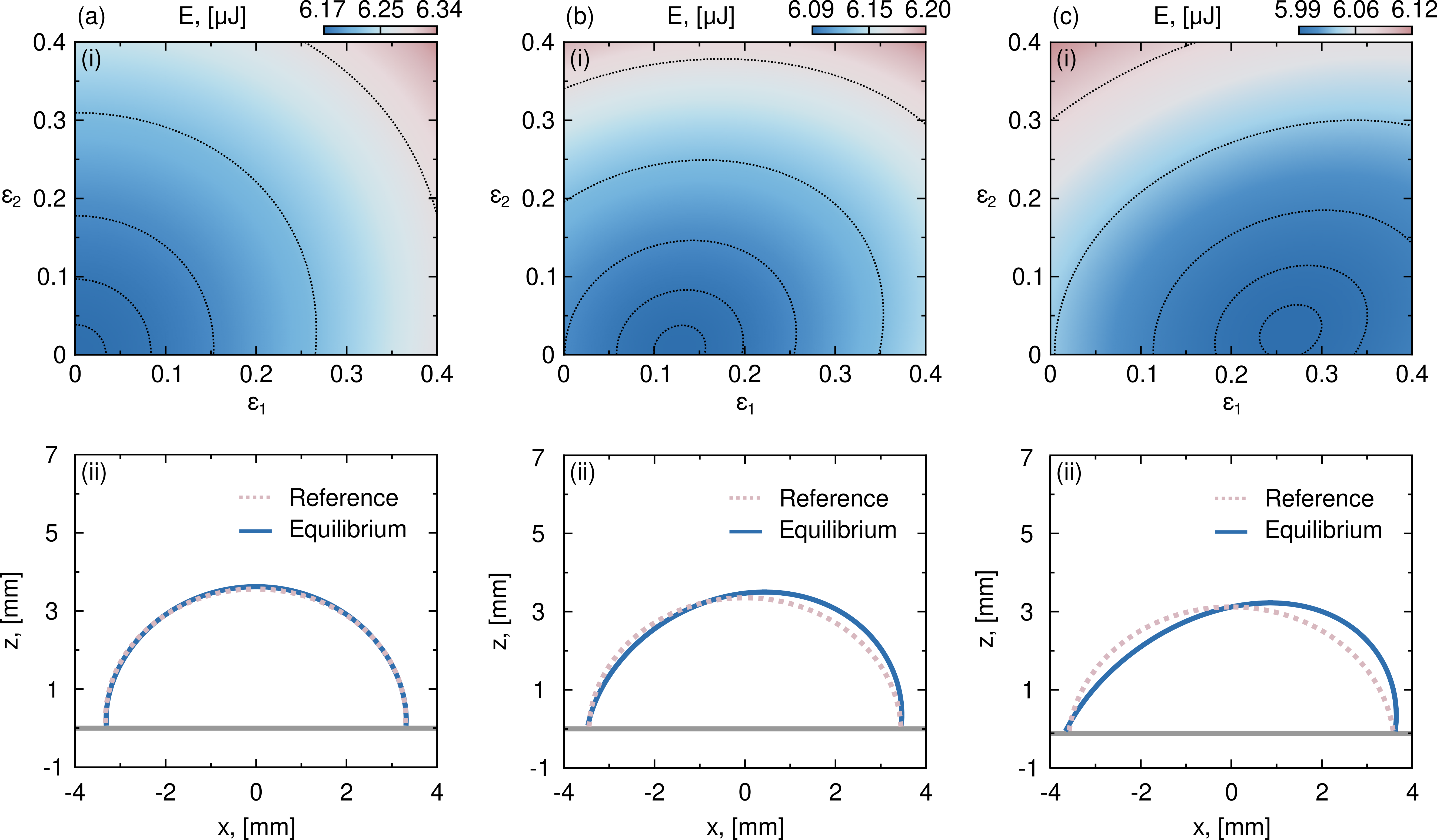}
\caption{\label{Fig.2} Representative free energy landscapes and equilibrium droplet profiles under uniform body force. An $80~\mu\mathrm{L}$ water droplet is considered on a smooth, inclined substrate under gravity, with a zero-forcing contact angle $\theta_0=92^\circ$. The gravity acceleration has components $a_x=g\sin\alpha$ and $a_z=-g\cos\alpha$, where $\alpha$ is the substrate inclination angle relative to the horizontal. Panels (a)--(c) correspond to $\alpha=0^\circ$, $10^\circ$, and $20^\circ$, respectively. In each panel, (i) shows the energy landscape $E(\varepsilon_1,\varepsilon_2)$, with configuration--independent wall energy and uniform pressure contributions omitted, and (ii) shows the $x$-$z$ profile selected at the energy minimum. Dashed and solid curves indicate the reference and equilibrium configurations, respectively. As $\alpha$ increases, the energy minimum shifts toward larger $\varepsilon_1^*$ and $\varepsilon_2^*$, accompanied by a progressively greater departure of the equilibrium profile from its reference configuration.}  
\end{figure*}

Once $\theta_0^*$ is determined, the spherical-cap reference configuration is constructed using the volume relation $V=(\pi/3)(r_c^*/\sin\theta_0^*)^3(2+\cos\theta_0^*)(1-\cos\theta_0^*)^2$. To describe the bulk response beyond the spherical-cap approximation, we introduce a two-parameter deformation manifold about the reference configuration. Specifically, the liquid--gas interface $A(\varepsilon_1,\varepsilon_2)$ is defined for $0\leq z\leq h$ by
\begin{equation}
A(\varepsilon_1,\varepsilon_2)
=
\left\{
\mathbf r:
\frac{(x-\varepsilon_1 z)^2}
{R_0^{*2}(\zeta)\Phi(\tilde{z})}
+
\frac{y^2}
{R_0^{*2}(\zeta)\Psi(\tilde{z})}
=1
\right\}.
\label{Eq.3}
\end{equation}
The dimensionless vertical coordinate is $\tilde{z}=z/h$, where the droplet height $h$ is governed by volume conservation, 
\begin{equation}
V=\pi h\int_0^1 R_0^{*2}(\zeta)
\sqrt{\Phi(\tilde{z})\Psi(\tilde{z})}\,\mathrm{d}\tilde{z}.
\label{Eq.4}
\end{equation}
Here, $R_0^*(\zeta)=
\left(
r_c^{*2}
-2r_c^*\zeta\cot\theta_0^*
-\zeta^2
\right)^{1/2}$ denotes the cross-sectional radius of the reference configuration evaluated at the mapped vertical coordinate $\zeta=h_0^{*}\tilde{z}$, where $h_0^*=r_c^*\tan(\theta_0^*/2)$ is the height of the reference configuration. The dimensionless variables $\varepsilon_1$ and $\varepsilon_2$ represent two distinct deformation modes: $\varepsilon_1$ shifts the center of each horizontal cross section according to $u(z)=\varepsilon_1z$, generating fore--aft asymmetry, whereas $\varepsilon_2$ controls the cross-sectional ellipticity through $\Phi(\tilde{z})=\exp[\varepsilon_2(1-\tilde{z})^m]$ and $\Psi(\tilde{z})=\exp[-\varepsilon_2(1-\tilde{z})^n]$. The exponents $m$ and $n$ specify how the two semiaxis modulations vary with height. Within the tested ranges, these exponents have only a minor influence on the predicted contact angles, supporting their treatment as fixed geometric parameters. We set $m=n=1.5$ throughout; see the Supplemental Material for sensitivity to these exponents.

Evaluating Eq.~\eqref{Eq.1} with \(p(\mathbf r)\) over the family of droplet geometries defined by Eqs.~\eqref{Eq.3} and~\eqref{Eq.4}, we obtain the free energy $E(\varepsilon_1,\varepsilon_2)$ for a droplet under a uniform body force. For a given applied forcing, Eq.~\eqref{Eq.2} fixes $\theta_0^{*}$, and hence the associated $\Delta\gamma$, over the deformation manifold. The wetted area is invariant under these deformations: at the substrate, $c=0$ and $u(0)=0$, while $\Phi(0)\Psi(0)=1$, giving
$\int_S \mathrm{d}S=\pi R_0^{*2}(0)$. Consequently, the wall free energy term contributes only a common offset to $E(\varepsilon_1,\varepsilon_2)$. Since $V$ is fixed, the uniform pressure contribution $-p_0V$ is also independent of $\varepsilon_1,\varepsilon_2$. Both terms shift only the free energy reference and can be omitted when locating the minimum. This simplification does not discard the interfacial effect, which is already incorporated through $\theta_0^*$ in defining the reference configuration. 

\section{Results and discussion}
After establishing $E(\varepsilon_1,\varepsilon_2)$, the equilibrium state at each prescribed forcing is selected by free energy minimization,
\begin{equation}
(\varepsilon_1^*,\varepsilon_2^*)
=
\operatorname*{arg\,min}_{\varepsilon_1,\varepsilon_2}
E(\varepsilon_1,\varepsilon_2).
\label{Eq.5}
\end{equation}
Combined with the volume constraint of Eq.~\eqref{Eq.4}, the minimizing pair $(\varepsilon_1^*,\varepsilon_2^*)$ specifies the complete equilibrium droplet geometry through Eq.~\eqref{Eq.3}. For concreteness, we illustrate the equilibrium selection in Eq.~\eqref{Eq.5} for an $80~\mu\mathrm{L}$ water droplet on an inclined substrate under gravity, as shown in Fig.~\ref{Fig.2}. Here, $a_x=g\sin\alpha$ and $a_z=-g\cos\alpha$, with $\alpha$ the substrate inclination angle relative to the horizontal. The parameters are $\sigma=0.072~\mathrm{N/m}$,
$\Delta\rho=1000~\mathrm{kg/m^3}$, $g=9.81~\mathrm{m/s^2}$, and $\theta_0=92^\circ$, with
$\alpha=0^\circ$, $10^\circ$, and $20^\circ$ in Figs.~\ref{Fig.2}(a)--\ref{Fig.2}(c), respectively. In each case, panel (i) shows the free energy landscape over the $(\varepsilon_1,\varepsilon_2)$ manifold, while panel (ii) presents the equilibrium $x$-$z$ profile associated with its minimum. At $\alpha=0^\circ$ [Fig.~\ref{Fig.2}(a)], the lateral pressure term, $\Delta\rho g\sin\alpha x$, vanishes, placing the energy minimum at the undeformed state; accordingly, the equilibrium profile coincides with the axisymmetric reference. Once the substrate is inclined [Fig.~\ref{Fig.2}(b)], $\Delta\rho g\sin\alpha x$ becomes finite, breaking the fore--aft symmetry of the pressure field and shifting the energy minimum to nonzero $\varepsilon_1^*$ and $\varepsilon_2^*$. The corresponding $x$--$z$ profile is asymmetric. Increasing the inclination further [Fig.~\ref{Fig.2}(c)], the minimum shifts toward larger $\varepsilon_1^*$ and $\varepsilon_2^*$ accompanied by a greater departure of the equilibrium interface from its reference.

\begin{figure}[t]
\includegraphics{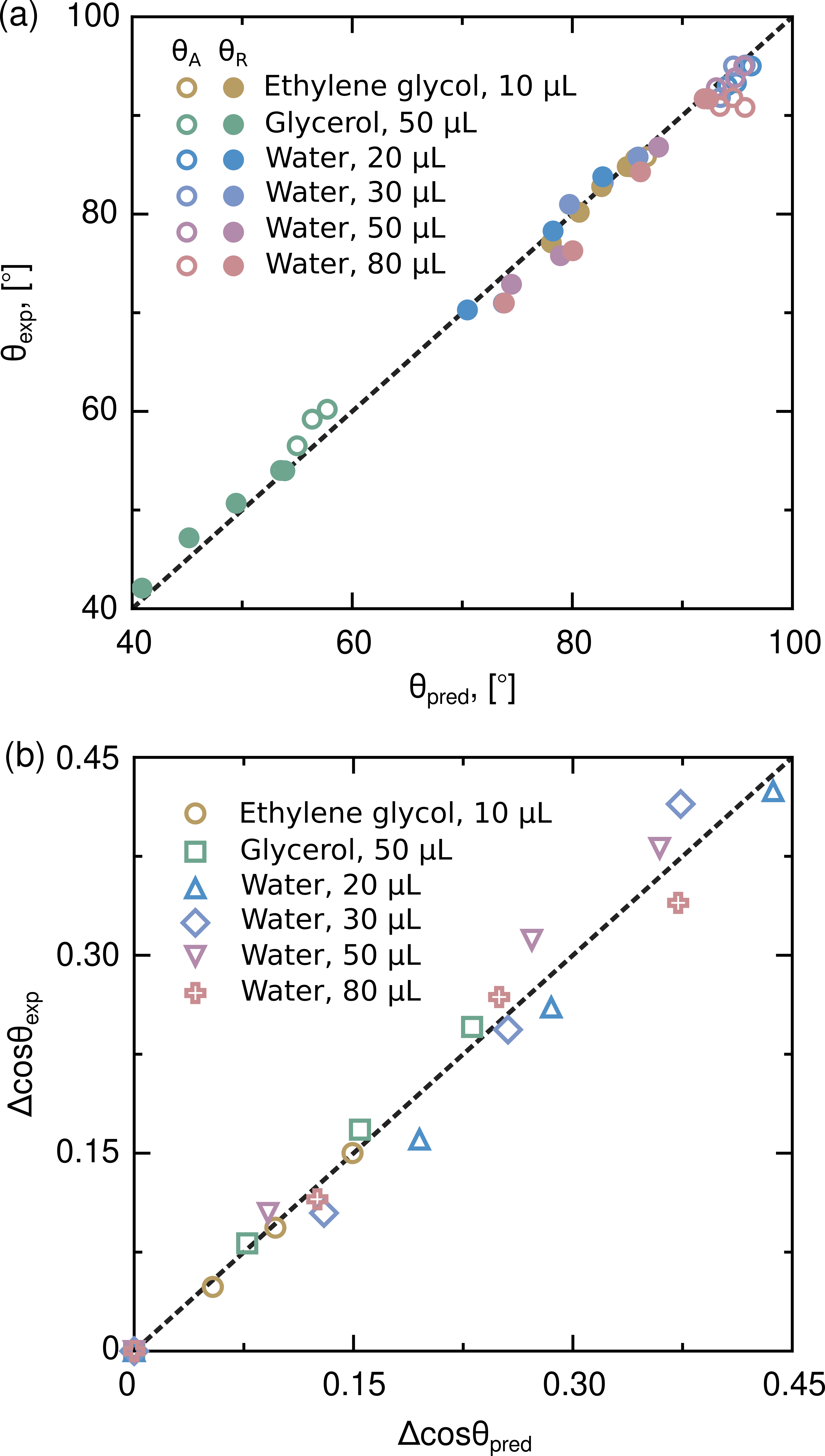}
\caption{\label{Fig.3} Quantitative validation of the predicted contact angles and the constitutive relation. (a) Predicted ($\theta_{\mathrm{pred}}$) versus measured ($\theta_{\mathrm{exp}}$) contact angles. Open and filled symbols represent $\theta_{\mathrm A}$ and $\theta_{\mathrm R}$, respectively. (b) Predicted ($\Delta \theta_{\mathrm{pred}}$) versus measured ($\Delta\theta_{\mathrm{exp}}$) contact angle asymmetry.
Predictions in (a) are obtained from Eq.~\eqref{Eq.7}, whereas those in (b) follow the constitutive relation in Eq.\eqref{Eq.9}. The measured asymmetry is calculated from the data in (a). Measurements for water droplets are taken from Ref.~\onlinecite{berejnov_effect_2007}, whereas those for ethylene glycol and glycerol are from Ref.~\onlinecite{bommer_depinning_2014}. }
\end{figure}

To characterize the fore--aft deformation of the equilibrium droplet, we consider the $x$-$z$ section at $y=0$,
\begin{equation}
x_{\pm}(z)
=
\varepsilon_1^* z
\pm
 R_0^*(\zeta)
\exp\!\left[
\frac{\varepsilon_2^*}{2}(1-\tilde{z})^m
\right],
\label{Eq.6}
\end{equation}
where $x_+$ and $x_-$ represent the branches on the $+x$ and $-x$ sides of the local centerline, respectively. Importantly, both branches arise from the same continuous equilibrium interface rather than from separately constructed profiles, thereby avoiding the artificial discontinuity associated with joining separate portions~\cite{tadmor_open_2021,tadmor_furmidge_2025}. Their local slopes at the substrate define the static apparent contact angles on the $+x$ and $-x$ sides, here denoted by $\theta_\mathrm{A}$ and $\theta_\mathrm{R}$, respectively. {For $m=n$, Eq.~\eqref{Eq.4} gives $h^*=h_0^*$, so that $\zeta=z$}; differentiating Eq.~\eqref{Eq.6} at the contact line and using $\left.\mathrm{d}R_0^*/\mathrm{d}z\right|_{z=0}
=-\cot\theta_0^*$ yields the following expressions for $\theta_\mathrm{A}$ and $\theta_\mathrm{R}$:
\begin{equation}
\begin{aligned}
\cot\theta_{\mathrm{A}}
&=
-\varepsilon_1^*
+
e^{\varepsilon_2^*/2}
\left[
\cot\theta_0^*
+
\frac{m\varepsilon_2^*R_0^*(0)}{2h^*}
\right];
\\
\cot\theta_{\mathrm{R}}
&=
\varepsilon_1^*
+
e^{\varepsilon_2^*/2}
\left[
\cot\theta_0^*
+
\frac{m\varepsilon_2^*R_0^*(0)}{2h^*}
\right].
\end{aligned}
\label{Eq.7}
\end{equation}
Equation~\eqref{Eq.7} provides a complete prediction of the static apparent contact angles for each prescribed forcing. In the undeformed limit, $\varepsilon_1^*=\varepsilon_2^*=0$, the two contact angles reduce to the reference value, $\theta_{\mathrm{A}}=\theta_{\mathrm{R}}=\theta_0^*$. Figure~\ref{Fig.3}(a) presents the predicted $\theta_{\mathrm{A}}$ and $\theta_{\mathrm{R}}$ together with the corresponding measurements for $10~\mu\mathrm{L}$ ethylene glycol, $50~\mu\mathrm{L}$ glycerol, and $20$, $30$, $50$, and $80~\mu\mathrm{L}$ water droplets. The $(\sigma,\Delta\rho)$ values are $(0.0477~\mathrm{N/m},1113~\mathrm{kg/m^3})$, $(0.063~\mathrm{N/m},1260~\mathrm{kg/m^3})$, and $(0.072~\mathrm{N/m},1000~\mathrm{kg/m^3})$, respectively. The agreement across different liquids, volumes, and forcing levels supports the reference state correction in Eq.~\eqref{Eq.2} and the equilibrium contact angle prediction of Eq.~\eqref{Eq.7}.

Next, we examine how the contact angle asymmetry responds to the applied forcing through the variation of the equilibrium state. Equation~\eqref{Eq.2} identifies $\Delta\rho a_x/\sigma$ as the relevant measure of the applied forcing. Letting $q\equiv\Delta\rho a_x/\sigma$, the free energy can be expressed as $E(\boldsymbol{\varepsilon},q)$, where $\boldsymbol{\varepsilon}=(\varepsilon_1,\varepsilon_2)^{\mathrm T}$. The equilibrium minimum satisfies $\nabla_{\boldsymbol{\varepsilon}}E(\boldsymbol{\varepsilon}^*,q)=\mathbf 0$. Differentiating this condition with respect to $q$ gives $\mathbf H
\mathrm d\boldsymbol{\varepsilon}^*/\mathrm dq +\mathbf b=\mathbf 0$, where $H_{ij}=\partial^2 E/\partial\varepsilon_i\partial\varepsilon_j$, $b_i=\partial^2 E/\partial\varepsilon_i\partial q$, {with $i,j=1,2$}, both evaluated at the equilibrium state. Here, $\mathbf H$ describes the local curvature of the free energy minimum, whereas $\mathbf b$ measures how $q$ perturbs the local free energy gradient. Since $\mathbf H$ is positive definite for the stable minima considered here, we obtain the equilibrium state response as $\mathrm d\boldsymbol{\varepsilon}^*/\mathrm dq = -\mathbf H^{-1}\mathbf b$. According to Eq.~\eqref{Eq.7}, the contact angle asymmetry,
$D\equiv\Delta\cos\theta=\cos\theta_\mathrm{R}-\cos\theta_\mathrm{A}$, can be regarded as $D(\boldsymbol{\varepsilon}^*,q)$. Differentiating $D(\boldsymbol{\varepsilon}^*,q)$ along the equilibrium branch and substituting the equilibrium state response above, we identify the
equilibrium contact angle asymmetry susceptibility $\chi$ as
\begin{equation}
\chi
\equiv
\frac{\mathrm dD}{\mathrm dq}
=
\frac{\partial D}{\partial q}
-
\left(\nabla_{\boldsymbol{\varepsilon}}D\right)^{\mathrm T}
\mathbf H^{-1}\mathbf b.
\label{Eq.8}
\end{equation}
The first term represents the explicit forcing dependence of $D$; the second captures the change in $D$ induced by the adjustment of the equilibrium state. In this form, $\chi$ distills the full equilibrium prediction into a single differential susceptibility, linking the variation of the contact angle asymmetry to the local free energy landscape and its perturbation by the applied forcing.

A further simplification follows from a first-order expansion of $D(q)$ about $q=0$, where $D(0)=0$ and $\chi_0\equiv\chi(0)$. The leading-order response is
\begin{equation}
D=\chi_0 q,
\qquad \text{i.e.,} \qquad
\Delta\cos\theta
=
\chi_0\frac{\Delta\rho a_x}{\sigma}.
\label{Eq.9}
\end{equation}
This equation converts the local differential response in Eq.~\eqref{Eq.8} into a finite-response constitutive relation: once $\chi_0$ is evaluated from Eq.~\eqref{Eq.8} at $q=0$, the static contact angle asymmetry follows directly from the applied forcing. To test Eq.~\eqref{Eq.9}, Fig.~\ref{Fig.3}(b) compares the predicted and experimental contact-angle asymmetries, $\Delta\cos\theta_{\rm pred}$ and $\Delta\cos\theta_{\rm exp}$, for the same six liquid--volume combinations examined in Fig.~\ref{Fig.3}(a). For each data point, $\Delta\cos\theta_{\rm pred}$ is evaluated from Eq.~\eqref{Eq.9} at the applied forcing, and $\Delta\cos\theta_{\rm exp}$ is determined from the measured $\theta_\mathrm{A}$ and $\theta_\mathrm{R}$. The data fall close to the diagonal $\Delta\cos\theta_{\rm exp}=\Delta\cos\theta_{\rm pred}$, demonstrating that Eq.~\eqref{Eq.9} serves as a finite-response constitutive law across different liquids and droplet volumes. For a known $\chi_0$, Eq.~\eqref{Eq.9} can be read in either direction, predicting the contact-angle asymmetry from the applied forcing or inferring the forcing from the measured asymmetry. Note that Eq.~\eqref{Eq.9} depends on the applied forcing only via $a_x$, so its constitutive form is not tied to a particular loading protocol; any loading protocol that produces a uniform body force can be cast in terms of $\Delta\rho a_x/\sigma$.

The significance of Eq.~\eqref{Eq.9} extends beyond its constitutive description. For a uniform body force, the total lateral force acting on the droplet is $F_x=\Delta\rho V a_x$. A static equilibrium configuration can persist only when this force is balanced by the lateral adhesion force $F_f$, such that $F_f=F_x$. Combining the force balance with Eq.~\eqref{Eq.9}, we obtain an equivalent force representation: $F_f= (V/\chi_0)\sigma\Delta\cos\theta$. From Eq.~\eqref{Eq.8}, $\chi_0$ has dimensions of area, making $V/\chi_0$ a quantity with dimensions of length. Comparison with the K--F relation, $F_{\mathrm{KF}}=k\omega\sigma\Delta\cos\theta$, 
gives $k\omega=V/\chi_0$,
and hence $k=V/(\chi_0\omega)$ when $\omega$ is taken as the droplet footprint width. Accordingly, $k\omega$ acquires a direct physical meaning as a response-derived length set by the equilibrium susceptibility, while $k$ measures this length relative to the footprint width.

\section{Conclusion}
In summary, we have proposed an alternative concept for linking static contact angle asymmetry to applied forcing. By incorporating the pressure field into both the interfacial and bulk contributions, we formulate the free energy of the forced droplet, whose minimization selects the equilibrium state and thereby determines the apparent contact angles and their asymmetry. Following the equilibrium branch as the forcing varies, we define an equilibrium susceptibility $\chi$. A first-order expansion of the asymmetry about zero forcing, with $\chi_0=\chi(0)$, then yields {a universal} constitutive relation
\begin{equation*}
\Delta\cos\theta
=
\chi_0\frac{\Delta\rho a_x}{\sigma}.
\end{equation*}
This relation places the connection between static contact angle asymmetry and applied forcing within a general equilibrium-response framework, in which the macroscopic asymmetry is governed by the free energy landscape and its susceptibility to applied forcing. This perspective may offer useful insight into static responses {in broader fluid--solid systems under forcing}.

L.~B. acknowledges financial support from the China Scholarship Council (CSC), File No.~202308120010. 
F.~W. acknowledges funding from the Helmholtz Association through the VirtMat project ``VirtMat P09: Wetting Phenomena,'' as part of the program ``MSE--Materials Science and Engineering,'' No.~43.31.01. 
R.Y. acknowledges funding from the Gottfried-Wilhelm Leibniz prize NE 822/31-1 of the German Research Foundation (DFG). 

\bibliography{references}

\end{document}